\documentstyle[aps,preprint]{revtex}
\begin{document}
\preprint{PAR-LPTHE/95-33, EHU-FT/9509, hep-th/95??}
\draft
\tighten
\hyphenation{Schwarz-schild}
\def\rs{r_\sigma}
\def\rt{r_\tau}
\def\rss{r_{\sigma\sigma}}
\def\rtt{r_{\tau\tau}}
\def\ts{t_\sigma}
\def\ta{t_\tau}
\def\tss{t_{\sigma\sigma}}
\def\taa{t_{\tau\tau}}
\def\ps{\phi_\sigma}
\def\pa{\phi_\tau}
\def\pss{\phi_{\sigma\sigma}}
\def\paa{\phi_{\tau\tau}}
\def\tes{\theta_\sigma}
\def\tea{\theta_\tau}
\def\tess{\theta_{\sigma\sigma}}
\def\teaa{\theta_{\tau\tau}}
\def\o{\over}
\def\s{\sigma}
\def\a{\alpha}
\title{Strings next to and inside  black holes}
\author{H. J. de Vega\cite{emLPTHE}}
\address{Laboratoire de Physique Th\'eorique et Hautes \'Energies
\\
Universit\'es Paris VI-VII - Laboratoire associ\'e au CNRS n$^{\rm o}$ 280
\\
Tour 16, 1er. \'et., 4, Place Jussieu, 75252 Paris Cedex 05 \\
FRANCE}
\author{I. L. Egusquiza\cite{emUPV}}
\address{Fisika Teorikoaren saila \\
Euskal Herriko Unibertsitatea\\
644 P.K. - 48080 BILBAO\\ SPAIN}
\date{\today}
\maketitle
\begin{abstract}
The string equations of motion and constraints are solved near the
horizon and near the  singularity of a Schwarzschild black hole.

In  a conformal gauge such that $\tau = 0$ ($\tau$ = worldsheet time
coordinate) corresponds to the
horizon ($r=1$) or to the  black hole  singularity ($r=0$), the
string coordinates express in power series in  $\tau$ near the horizon
and  in power series in $\tau^{1/5}$ around $r=0$.

We compute  the string invariant size and the string
energy-momentum tensor. Near the horizon both are finite and analytic.
Near the  black hole  singularity, the string size, the string energy
and the transverse pressures (in the angular directions) tend to
infinity as $r^{-1}$.
To leading order near  $r=0$, the string behaves as two dimensional
radiation. This  two spatial dimensions are describing the $S^2$ sphere in the
Schwarzschild manifold.
\end{abstract}
\pacs{11.25.-w,04.20.Dw, 04.70.-s, 11.27+d  }
\section{Introduction and Motivations}

The systematic investigation of strings   in curved spacetimes
started in \cite{plb,negro} has uncovered a variety of new physical phenomena
(see \cite{erice} for a review). These results  are relevant both for
fundamental (quantum) strings and for cosmic strings, which behave in
an  essentially classical  way \cite{twbk}.

Black-hole spacetimes are probably the most relevant backgrounds
to study strings
propagating into a spacetime singularity (see  \cite{ondp}
for plane wave null singularities). Strings in Schwarzschild
black holes have been previously investigated in
refs.\cite{negro,fr,cl,hjile,ls,al}. Different kinds of symmetric Ans\"atze
that {\bf separate} the variables $\s$ and $\tau$ have been
proposed. In this way the string evolution equations (which are
non-linear partial  differential  equations) become ordinary differential
equations.

In ref.\cite{fr} stationary solutions are obtained with the spatial
variables depending only on   $\s$ and the temporal variable  set as $t
= \tau$. In this way
infinitely long  string solutions outside the horizon were constructed. String
solutions inside the horizon are obtained in ref.\cite{ls} by
exchanging $\s \leftrightarrow \tau$ in the previous Ansatz. These string are
also infinitely long. All solutions are expressed in terms of elliptic
functions.

Ring solutions are obtained in ref.\cite{hjile} through the Ansatz
$\phi = n \s$ and assuming the other coordinates to depend only on
$\tau$. These solutions describe strings that can propagate from
infinity. As they approach the black hole, they may fall or not into the
singularity depending on the initial conditions. A special case
describes strings oscillating inside the horizon \cite{hjile,al}. This
special solution expresses  in terms of elementary functions.

We construct in this paper {\bf generic} solutions of the string
equations of motion and constraints in Schwarzschild black holes. We do
that near the horizon and near the black hole singularity.
We consider here closed strings. The generalization to open strings is
straightforward.

Through appropriate conformal transformations on the world-sheet, we
map in the generic case
the intersection of  the string world-sheet with the horizon into
the line $\tau = 0$. A similar transformation can be performed for
the black hole singularity $r = 0$. We then study the   string
equations of motion and constraints by expanding in $\tau$ around
$\tau = 0$. Near the black hole singularity, it turns out that the string
solutions possess a  series expansion in integer powers of $\tau^{1/5}$.
Around the horizon, strings solutions are analytic in $\tau$ and only
integer powers of  $\tau$ appear.

Calling upon  the conformal freedom, our solutions depend on four
arbitrary functions of  $\s$. These functions can be expressed in
terms of the initial data. We thus see that only the transverse coordinates
are physical degrees of freedom.

In Kruskal-Szekeres coordinates $u,v,\theta,\phi$ [see
eq.(\ref{schkru})], a generic string solution behaves near the
black hole singularity as follows:

\begin{eqnarray}\label{solint}
u(\s,\tau)&=& e^{a(\s)} \left\{ 1\right. -    {1\o 4} \, \gamma(\s)^4
\; \tau^{ 4/5}  \left[ 1 + O(\tau^{2/5}) \right] \cr \cr
&-& \left.  \gamma(\s)^6 \;
{{f'(\sigma)\nu(\sigma)\sin^2g(\sigma)+\mu(\sigma)}\over
{28\;  a'(\sigma)}}\; \tau^{7/5} \left[ 1 + O(\tau^{2/5}) \right] \right\}
\; , \cr \cr
v(\s,\tau)&=& e^{-a(\s)} \left\{ 1\right. -   {1\o 4} \, \gamma(\s)^4
\; \tau^{ 4/5}  \left[ 1 + O(\tau^{2/5}) \right] \cr \cr
&+& \left.  \gamma(\s)^6 \;
{{f'(\sigma)\nu(\sigma)\sin^2g(\sigma)+\mu(\sigma)}\over
{28\;  a'(\sigma)}}\; \tau^{7/5} \left[ 1 + O(\tau^{2/5}) \right] \right\}
\; , \cr \cr
\theta(\s,\tau)=  g(\s) &+&   \tau^{1/5}\; \mu(\s) + O(\tau^{3/5}) , \cr \cr
\phi(\s,\tau)=   f(\s) &+&  \tau^{1/5}\; \nu(\s) + O(\tau^{3/5}) .
\end{eqnarray}
where
\begin{equation}
  [2 \,  \gamma(\s)]^2 = \mu(\s)^2 +  \nu(\s)^2 \; \sin^2 g(\s) \; .
\end{equation}
In addition, the (time-like) coordinate $r$ vanishes as,
\begin{equation}
r(\s,\tau) =\gamma(\s)^2 \;  \tau^{2/5} +   O(\tau^{4/5})\; .
\end{equation}
Notice that the angular coordinates $\theta, \phi$ vary with $\tau$
near $\tau =0, r=0$ faster than the $u,v$ or $r$ coordinates.

The string behaviour near the horizon is analytic in  $\tau$. It is
given in sec.III, eqs.(\ref{domhor}-\ref{tetag}).

We then compute the string size and energy-momentum near the
black hole singularity.  The invariant size $S$ \cite{erice} grows as
$r^{-1}$ for $r\to 0$,
\begin{equation}
S =  {{ 4\,  {a'(\s)}^2}\o r} + O(1)
\end{equation}
The string infinitely stretches when it falls into the   $r=0$ singularity.
This is due to the infinitely growing gravitational forces that act
then on the string.  The string stretching near  $r=0$ was observed in
ref.\cite{cl} using perturbative methods.

The string energy diverges also as $r^{-1}$ for $r\to 0$,
\begin{equation}
E = { 1 \o {2\pi\alpha'}}\;
{2\o {5\;r}}\int_0^{2\pi}{\rm d}\s\,|\gamma(\s)|^5 \, + O(1) \to +\infty \; ,
\end{equation}
where $2\pi\alpha'$ stands for the inverse string tension.

The pressure in the angular directions  $\theta, \phi$ tends to infinity
with the same rate:
\begin{eqnarray}
P_{\phi}  &=&  { 1 \o {2\pi\alpha'}}\;
 {1\o {10\;r}}
\int_0^{2\pi}{\rm d}\s\, \nu(\s)^2 \;  \sin^2 g(\s) \;  |\gamma(\s)|^3
\,\to +\infty\, ,\cr\cr
P_{\theta} &=&  { 1 \o {2\pi\alpha'}}\;
{1\o {10\;r}}
\int_0^{2\pi}{\rm d}\s\, \mu(\s)^2 \; |\gamma(\s)|^3
\,\to +\infty .
\end{eqnarray}
Thus, to leading order,
$$
E = P_{\theta}+ P_{\phi} \quad \mbox{for}\quad r \to 0 \; .
$$
which is the  behaviour of a two-dimensional ultrarelativistic gas.
 This relation means that the string behaves to leading order as  {\bf
two}-dimensional massless
particles. This is the so-called dual to unstable behaviour \cite{erice,backr}
(here for two spatial dimension).

As is well known, the Schwarzschild manifold has a $M \bigotimes S^2$
structure, where $M$ stands for the part described by the $u,v$ coordinates.
The present results indicate
that test strings near the black hole singularity behave as massless
particles propagating in the compact $ S^2 $  manifold.

Near the horizon both the string size and the energy-momentum tensor
are finite and analytic in Kruskal-Szekeres coordinates. The trace of the
energy-momentum tensor is positive at the horizon.

It must be noticed that the resolution method used here for strings in
black hole spacetimes is analogous  the expansions for $\tau\to 0$
used in ref.\cite{sv,gsv} for  strings in cosmological spacetimes.

For strings at large distances from the black hole where the
gravitational field is weak, the string propagation can be solved by
perturbing the Minkowski solutions. Actually, one can take as zeroth
order solution the center of mass motion, $ \; q^A + 2 p^A \a' \tau \;
$ as it has been done  in ref.\cite{negro}.

In summary, the exact string behaviour near the black hole singularity
and near the  black hole horizon is presented in this paper.

\section{Equations of motion.}

The Schwarzschild metric in Schwarzschild coordinates
$(t,r,\theta,\phi)$ takes the following form:
\begin{equation}\label{schsch}
\displaystyle{
ds^2=-\left(1-{1\o{r}}\right){\rm d}t^2 + {{{\rm
d}r^2}\o{1-{1\o r}}} + r^2({\rm
d}\theta^2+\sin^2\theta\>{\rm d}\phi^2)\,, }
\end{equation}
where we choose units where  the Schwarzschild radius
$R_s=2m = 1$.

Since we are interested in the whole Schwarzschild manifold and not
just in the external part $r > 1$ where the static Schwarzschild
coordinates are
appropriate, we consider the Kruskal-Szekeres coordinates
$(u,v,\theta,\phi)$ defined by
\begin{equation}\label{dfkru}
u= t_K - r_K \equiv \sqrt{1-r}\; e^{(r-t)/2} \quad , \quad
v= t_K + r_K \equiv \sqrt{1-r}\; e^{(r+t)/2} \; .
\end{equation}
for $v \geq  0, u \geq 0$ and by
\begin{equation}\label{dfkru2}
u= t_K - r_K \equiv -\sqrt{r-1}\; e^{(r-t)/2} \quad , \quad
v= t_K + r_K \equiv \sqrt{r-1}\; e^{(r+t)/2} \; .
\end{equation}
for $v  \geq 0, u \leq 0$. For $v \leq 0$ one just flips the sign of
$v$ in eq.(\ref{dfkru}) or (\ref{dfkru2}) \cite{bibneg}.

The coordinate $ t_K $ is a time-like coordinate, and $ r_K $ spacelike.
In  Kruskal-Szekeres coordinates the Schwarzschild metric takes the form,
\begin{equation}\label{schkru}
ds^2=-{4\over{r}}~e^{-r}\;du\,dv
+r^2~({\rm d}\theta^2+\sin^2\theta\>{\rm d}\phi^2)\,.
\end{equation}
$r$ is a function of the product $uv$ defined by the inverse function
of
$$
uv = [1-r] \, e^r \; .
$$
for $u v \geq 0$. The metric is such coordinates is  regular
everywhere except at its singularity,  $r=0$.

The string equations of motion  in Schwarzschild
coordinates and in the conformal gauge, are
\begin{eqnarray}\label{eqsch}
\rs\ts - \rt\ta + r(r-1)(\tss-\taa)&=&0\,,\cr
{{2r}\over{1-r}}(\rtt-\rss) - {1\over{r^2}}(\ta^2-\ts^2) + 2r ( \tea^2
-\tes^2)&+&\quad\cr
\displaystyle{
2\,r\sin^2\theta\>(\pa^2-\ps^2)} \displaystyle{\,+\,
{1\over{(r-1)^2}}(\rt^2-\rs^2)}&=&0\,.
\end{eqnarray}
\begin{eqnarray}\label{eqang}
r\sin\theta\>(\paa-\pss) + 2\, r\cos\theta\>(\pa\tea-\ps\tes) + 2 \,\sin\theta
\, (\rt\pa-&\rs\ps)=0\,,\cr \cr
r\,(\teaa-\tess)+ 2(\rt\tea-\rs\tes) - r\sin\theta\>\cos\theta\>
(\pa^2-\ps^2)&=0\, .
\end{eqnarray}
The constraints  in Schwarzschild coordinates are
\begin{eqnarray}\label{vinsch}
{{1-r}\over{r}}(\ts^2+\ta^2) + {r\over{r-1}}(\rt^2+\rs^2)
&+& r^2 \left[\tea^2+\tes^2 + \sin^2\theta\>(\pa^2+\ps^2)\right]=0\,,\cr
{{1-r}\over{r}}\ta\ts +  {r\over{r-1}} \rt\rs &+ & r^2\left(\tea\tes
+ \sin^2\theta\>\pa\ps\right)=0\,.
\end{eqnarray}

The string equations of motion in  Kruskal-Szekeres coordinates take
the form (always  in the conformal gauge),
\begin{eqnarray}\label{eqks}
u_{\tau \tau} -u_{\sigma\sigma} +{1 \o r}\left( 1 + {1 \o r}
\right)\; e^{-r}\; v\; \left[(u_{\tau})^2 -(u_{\sigma})^2\right]
- {{r\, u}\o 2}  \left[  \tea^2 -\tes^2
+\sin^2\theta\>(\pa^2-\ps^2) \right] = 0 \cr \cr
v_{\tau \tau} -v_{\sigma\sigma} +{1 \o r}\left( 1 + {1 \o r}
\right)\; e^{-r}\; u\; \left[(v_{\tau})^2 -(v_{\sigma})^2\right]
- {{r\, v}\o 2}  \left[  \tea^2 -\tes^2
+\sin^2\theta\>(\pa^2-\ps^2) \right] = 0 \; ,
\end{eqnarray}
plus eqs.(\ref{eqang}) for the angular coordinates.

The constraints  in Kruskal-Szekeres coordinates are
\begin{eqnarray}\label{vinks}
-{4\over{r}}~e^{-r}\,
\left(u_{\sigma}\;v_{\sigma}+u_{\tau}\;v_{\tau}\right) +
 r^2 \left[\tea^2+\tes^2 + \sin^2\theta\>(\pa^2+\ps^2)\right]=0\,,\cr
-{4\over{r}}~e^{-r}\,
\left(u_{\tau}\;v_{\sigma}+u_{\sigma}\;v_{\tau}\right) +
 r^2\left(\tea\tes + \sin^2\theta\>\pa\ps\right)=0 \, .
\end{eqnarray}
Notice that both the equations of motion and constraints are invariant
under the exchange $u \leftrightarrow v$.

Also notice that  the equations of motion and constraints in Kruskal-Szekeres
coordinates are regular everywhere except at the singularity $r=0$.

We shall consider closed strings where the string coordinates must be
periodic functions of $\s$:
 \begin{equation}\label{perio}
u(\s+2\pi,\tau) = u(\s,\tau) \; , \; v(\s+2\pi,\tau)= v(\s,\tau) .
\end{equation}
Therefore, the  angular coordinates $\theta,\phi$ may be just
quasiperiodic functions of $\s$:
\begin{equation}\label{qperio}
\theta(\s+2\pi,\tau) = \theta(\s,\tau) + \mbox{mod}\,2\pi\; , \;
\phi(\s+2\pi,\tau) = \phi(\s,\tau) + 2 n \pi ,
\end{equation}
where $ n$ is an integer.

\section{ Strings Across the Horizon}

Let us study the string behaviour in the region near the horizon.
The  Kruskal-Szekeres coordinates are the appropriate ones here.
The horizon is characterized by $uv=0$. Let us consider a string
crossing the right part of it, that is $u=0, v > 0$.

We can write the curve describing the intersection of the horizon with
the world-sheet as
\begin{equation}\label{hojhor}
x_+ = \chi(x_-),
\end{equation}
whenever this intersection is nondegenerate. (Here $ x_{\pm} \equiv
\tau \pm \s $).

Upon a conformal transformation,
\begin{equation}\label{confg}
x_+ \to x_+' = f(x_+) \quad , \quad x_- \to x_-' = g(x_-),
\end{equation}
we can map the curve (\ref{hojhor}) into $\tau' = 0$ by an appropriate
choice of $f$ and $g$. For example, we can choose
$$
 f(x_+) = x_+  \quad , \quad g(x_-) = -  \chi(x_-).
$$
This defines our choice of gauge. From now on, we rename $\tau'$ and $\s'$
by $\tau$ and $\s$, respectively. Notice that this choice does not
completely fix the gauge. We can still perform transformations that
leave the line $\tau = 0$ unchanged. This is the case for the
following class of conformal mappings
\begin{equation}\label{diago}
x_+ \to x_+' = \varphi(x_+) \quad , \quad x_- \to x_-' = -\varphi(-x_-),
\end{equation}
where $\varphi(x)$ is an arbitrary function. Eq. (\ref{diago}) can be
written as,
\begin{eqnarray}\label{diag}
\tau' &=& {1 \o 2} \left[ \varphi(\tau + \s) - \varphi(\s - \tau)
\right] = \tau \, \varphi'(\s)\, + {1 \o 6}\, \tau^3 \;
\varphi'''(\s)+ O(\tau^4) \cr \cr
\s' &=& {1 \o 2} \left[ \varphi(\tau + \s) + \varphi(\s - \tau)
\right] = \varphi(\s)\, + {1 \o 2}\, \tau^2 \;
\varphi''(\s)+ O(\tau^4)
\end{eqnarray}
The transformations  (\ref{diago}) represent a diagonal subgroup of
the set of left-right conformal transformations (\ref{confg}).

\subsection{Equatorial Solutions}

Let us first  investigate strings on the plane $\theta = \pi/2$.
This restriction is compatible with the equations of motion and
constraints and it means that the string moves in an equatorial plane.

We can then propose the following expansion near $v=0$,
\begin{eqnarray}\label{domhor}
v(\s,\tau) &=& e \left[ \tau \,  \, c_1(\s) + \tau^2 \, c_2(\s) +
O(\tau^3) \right] \cr \cr
u(\s,\tau) &=& b_0(\s) + \tau \, b_1(\s) + \tau^2 \, b_2(\s) + O(\tau^3)
\cr \cr
\phi(\s,\tau) &=& p_0(\s)  + \tau \, p_1(\s) + \tau^2 \, p_2(\s) +
O(\tau^3) \; ,
\end{eqnarray}
where $e = 2.718281828459 \ldots $ . Inserting
 eqs.(\ref{domhor}) in  eqs.(\ref{eqang},\ref{eqks}) and constraints
(\ref{vinks}) yields
\begin{eqnarray}\label{domhor2}
c_2(\s) &=& -  b_0(\s) \, c_1(\s)^2 \; , \cr \cr
b_1(\s) &=& {{ c_1(\s) \,  [b_0'(\s)]^2}\o{ [p_0'(\s)]^2}}
+ {{ [p_0'(\s)]^2}\o {4 \,  c_1(\s)}}  \; ,\cr \cr
 p_1(\s) &=& {{2  c_1(\s)\, b_0'(\s)}\o {p_0'(\s)}} \; .
\end{eqnarray}
In addition $r(\s,\tau)$ is given by the expansion
\begin{equation}\label{rhor}
r(\s,\tau) = 1 -  b_0(\s) \, c_1(\s) \; \tau -\left\{ {{ [c_1(\s) \,
b_0'(\s)]^2}\o{ [p_0'(\s)]^2}} + {1 \o 4} \,  [p_0'(\s)]^2
\right\}\tau^2 + O(\tau^3) \; .
\end{equation}
The expansion can be pushed to arbitrary high orders in $\tau$
yielding the series coefficients in terms of the arbitrary functions
$ b_0(\s),  c_1(\s)$ and
$p_0(\s)$. We have easily generated a number of terms with the help
of Mathematica. The number of arbitrary functions can be reduced to
two using the conformal freedom (\ref{diag}).

The boundary conditions (\ref{perio}-\ref{qperio}) require here:
\begin{equation}\label{concor}
 b_0(\s+ 2\pi ) =  b_0(\s)\; , \; c_1(\s + 2\pi )= c_1(\s)\; , \;
p_0(\s+ 2\pi)=p_0(\s) +2 n \pi \; .
\end{equation}

\subsection{General String Solutions}

Let us consider now a general string solution near the horizon
$r=1$. We keep working in a gauge such that $\tau = 0$ corresponds to
$r=1$.

Eqs. (\ref{domhor}) are then supplemented by
\begin{equation}\label{tetah}
\theta(\s,\tau)= t_0(\s) + t_1(\s) \, \tau + t_2(\s) \, \tau^2 +
O(\tau^3)\; .
\end{equation}
Inserting eqs.(\ref{domhor}, \ref{tetah}) into
eqs.(\ref{eqang},\ref{eqks}) and constraints (\ref{vinks}) yields
\begin{eqnarray}
u(\s,\tau) & = &
b_0 +\nonumber\\
& + &{\tau \o {4 c_1}}  \Bigg[(p_1 + p_0')^2 \, \sin^2 t_0 +
( t_1 + t_0')^2 \Bigg] +\nonumber\\
& + & {{\tau^2}\o 4}
\Bigg[b_0\, ({p_1}^2 -{p_0'}^2)\,\sin^2 t_0 +
  b_0\,({t_1}^2 - {t_0'}^2) + 2 \, b_0''
\Bigg] +O(\tau^3)\; ,\nonumber\\
v(\sigma,\tau) &=& e \left\{ \,c_1\tau
 - \,b_0\,{{c_1}^2}  \; \tau^2 \right. \nonumber\\
& + &   \,{{\tau^3}\o 12}\,
\Bigg[6 \, {b_0}^2 \, {c_1}^3 + 4 \, {c_1}^2 \,b_0' + 2 \, c_1''-
  c_1\,p_0'\,( p_1 + p_0')  \nonumber\\
 & - & \left.    2\, c_1\, t_0'(t_0'+ t_1)
+ \cos (2\,t_0)\,c_1\,p_0'(p_1+p_0')\Bigg] + O(\tau^4)\right\} \; ,\nonumber\\
r(\s,\tau)&=&1
- b_0\,c_1 \;  \tau+\nonumber\\
& - &  {  \tau^2 \o 4} \; \Bigg[
 \; \sin^2t_0 \; (p_0' + p_1)^2 + \; (t_0' +
t_1)^2  c_1\,b_0'  \Bigg]+O(\tau^3)\; , \nonumber\\
\phi(\s,\tau)&=& p_0 + p_1\tau + \left[
b_0\,c_1\,p_1 -   ( \,p_1\,t_1 -   \,p_0'\,t_0')  \cot t_0 +
  {{p_0''}\over 2}\right]\; \tau^2 +O(\tau^3)\; ,  \nonumber\\
\label{tetag}
\theta(\s,\tau)&=& t_0 + t_1\tau +\left[
{{\sin (2\,t_0)}\over 4} ({p_1}^2 - {p_0'}^2)
+ b_0\,c_1\,t_1 +
  {{t_0''}\over 2}\right]\; \tau^2 +O(\tau^3) \; . \;  \nonumber\\
\end{eqnarray}
Here,
\begin{equation}
 p_1(\s)= {{2 \, c_1 b_0' - t_1 t_0'}\over{p_0' \; \sin^2 t_0}}\; .
\end{equation}

We have here five arbitrary functions $b_0(\s), c_1(\s),  p_0(\s),
t_0(\s)$ and  $t_1(\s)$. This number can be reduced to four
using the diagonal conformal transformations (\ref{diag}).

\section{ Strings Near the Singularity $r=0$}

Let us consider the solution of eqs.(\ref{eqang},\ref{eqks}) and constraints
(\ref{vinks}), near $r=0$. That is to say, near $uv=1$.

For a generic world-sheet, we choose the gauge such that   $\tau=0$
corresponds to the string at the singularity $uv=1$. This can be
achieved as shown  in sec. III for the horizon.

Notice that we use in sec. III and IV the same symbols $(\s,\tau)$ for
different  world-sheet coordinates. In sec. III,   $\tau=0$
corresponds to  $r=1$, whereas in sec. IV, $\tau=0$ corresponds to  $r=0$.

\subsection{Equatorial Solutions}

Let us begin to investigate strings on the plane $\theta = \pi/2$.
This restriction is compatible with the equations of motion and
constraints and it means that the string moves in an equatorial plane.

Near the singularity $uv=1$, we propose for $ \tau\to 0 $ the Ansatz
\begin{eqnarray}\label{domi}
u(\s,\tau)&=& e^{a(\s)} \left[ 1 - \tau^{\a} \, \beta(\s) + \ldots
\right]\cr\cr
v(\s,\tau)&=& e^{-a(\s)} \left[ 1 - \tau^{\a'} \, {\hat \beta}(\s) + \ldots
\right]\cr\cr
\phi(\s,\tau)&=& f(\s) + 2\, \gamma(\s)\; \tau^{\lambda}  + \ldots \; .
\end{eqnarray}
[The factor $2$ is there just for future convenience].

Inserting eqs.(\ref{domi}) in  eqs.(\ref{eqang},\ref{eqks}) and constraints
(\ref{vinks}) yields
\begin{eqnarray}\label{expo}
\a = \a'= 4/5 \quad , \quad \lambda = 1/5 ,\cr
 {\hat \beta}(\s) =  \beta(\s) \quad , \quad  \beta(\s) = {1 \o 4}\,
\gamma(\s)^4 .
\end{eqnarray}
The functions $a(\s), f(\s) $ and $ \gamma(\s)$ are arbitrary.

The coordinate $r$ then vanishes as
\begin{equation}
r(\s,\tau) =\gamma(\s)^2 \;  \tau^{2/5} + \ldots \; .
\end{equation}

We have pushed the resolution of  eqs.(\ref{eqang},\ref{eqks}) and constraints
(\ref{vinks}) to higher orders with the help of Mathematica. We find
that the corrections to the leading behaviour appear as integer
powers of $\tau^{2/5}$. That is,  terms in $\tau^{6/5}, \tau^{8/5},
\tau^{10/5}, \ldots$
in $u$ and in $v$, and terms in  $\tau^{3/5}, \tau^{5/5},
\tau^{7/5},\ldots$ in $\phi$.
The subdominant contributions start with the  order $\tau^{7/5}$
in  $u$ and in $v$. The  piece in  $\phi$ of order  $\tau^0$  is
directly connected with the subdominant contributions in  $u$ and in $v$.

We find,
\begin{eqnarray}\label{uvfi0}
u(\s,\tau)&=& e^{a(\s)} \left\{ 1\right. -\left. {1\o 4} \, \gamma(\s)^4 \;
\tau^{ 4/5}
-{1 \o 21}\;  \left( 2 \, \gamma(\s)^6 + {75 \o 2}\,
{{a'(\s)^2}\o{\gamma(\s)^4}} \right) \, \tau^{6/5} \;  \left[1 +  O(
\tau^{ 2/5})  \right] \right.\cr
&-& \left.{{\gamma(\s)^7 \, f'(\s)}\o {14 \, a'(\s)}}\; \tau^{7/5}
 \left[1 +  O(\tau^{ 2/5})  \right]\right\} , \cr \cr
v(\s,\tau)&=& e^{-a(\s)} \left\{ 1\right. -\left. {1\o 4} \,
\gamma(\s)^4 \; \tau^{ 4/5}
-{1 \o 21}\;  \left( 2 \, \gamma(\s)^6 + {75 \o 2}\,
{{a'(\s)^2}\o{\gamma(\s)^4}} \right) \, \tau^{6/5} \;  \left[1 +  O(
\tau^{ 2/5})  \right] \right.\cr
&+& \left.{{\gamma(\s)^7 \, f'(\s)}\o {14  \, a'(\s)}}\; \tau^{7/5}
 \left[1 +  O(\tau^{ 2/5})  \right]\right\} , \cr \cr
\phi(\s,\tau)&=&   f(\s) + {5 \o 21} \;  \left( {{4\,  f'(\s) \,  a''(\s)}\o{
7\, a'(\s)}} + {{4  \,  f'(\s)  \, \gamma'(\s)}\o{3\, \gamma(\s)}} + {53 \o
42} \; f''(\s) \right) \; \tau^2 +  O(\tau^{ 12/5}) \cr &+&
2 \, \gamma(\s)  \, \tau^{1/5}  \left\{ 1 +
{ 2 \o{21\, \gamma(\s)^8 }}\,  \left[ \gamma(\s)^{10} - 25 \,
a'(\s)^2\right]\tau^{ 2/5}  +  O(\tau^{ 4/5}) \right\} \; ,
\end{eqnarray}
and
\begin{eqnarray}\label{r0}
r(\s,\tau) &=&  \gamma(\s)^2 \;  \tau^{2/5} \left[ 1 +
{{\tau^{2/5}}\o{7}}\left(- \gamma(\s)^2
+{{25\, a'(\s)^2}\o{\gamma(\s)^6}} \right) + O( \tau^{ 4/5})   \right] \cr \cr
&+&{25 \o 441}\, \tau^{ 11/5} \, \left[ 35 \, \gamma'(\s)\,  f'(\s) -
{{6 \,f'(\s)\, \gamma(\s)
\,a''(\s)} \o { a'(\s)}} -  \gamma(\s) \, f''(\s)  \;
  +  O(\tau^{ 2/5})  \right] \; .
\end{eqnarray}
The functions  $ f(\s), a(\s)$ and $ \gamma(\s)$ are
arbitrary and can be expressed in terms of the  initial data.
In the Appendix we give $u,v,\phi,r$ up to the order $\tau^{ 16/5} $.

Both the equations of motion and constraints are invariant
under the exchange $u \leftrightarrow v$ but not the boundary
conditions at $\tau = 0$. They differ by  $ a(\s) \leftrightarrow
-a(\s)$ as we see from eqs.(\ref{domi}). Therefore one can obtain
$u(\s,\tau)$ from  $v(\s,\tau)$ and
viceversa just by flipping the sign of  $ a(\s) $.

Notice that  $u/v$ is $\tau$ independent up to order $\tau^{7/5}$.
Since $u/v = e^{-t}$, this imply that the spatial coordinate $t$ is
only $\s$-dependent up to $O(\tau^{7/5})$. More precisely,
\begin{equation}
t(\s,\tau) = \log{v\o u} = - 2 \, a(\s) + {{  \gamma(\s)^7 \,  f'(\s)}\o {
7 \, a'(\s)}}\; \tau^{7/5}  +   O(\tau^{ 9/5}) \; .
\end{equation}
In other words, $t(\s,\tau)$ varies slower than the other coordinates
$\phi$ and $r$ when the string approaches the black hole singularity
($\tau \to 0$).

The boundary conditions (\ref{perio}-\ref{qperio}) require here:
\begin{equation}
 a(\s+ 2\pi ) =  a(\s)\; , \; \gamma(\s + 2\pi )= \gamma(\s)\; , \;
f(\s+ 2\pi)=f(\s) +2 n \pi \; .
\end{equation}

Recall that we can still perform diagonal conformal transformations
(\ref{diago}). It is easy to check that eqs.(\ref{uvfi0}) keep their
form under the transformations (\ref{diag}) when we expand in powers
of $\tau$.

We can use the arbitrary function $\varphi(\s)$ to fix one function among
 $ f(\s), a(\s)$ and $ \gamma(\s)$.
For example, we can set to zero the $O(\tau^2)$ part in
$\phi(\s,\tau)$ [see eq.(\ref{uvfi0})].
We are thus left with {\bf two} arbitrary functions. They describe
precisely the transverse degrees of freedom of the equatorial string.

\subsection{General String Solutions}

Let us consider now a general string solution near the singularity
$r=0$. We keep working in a gauge such that $\tau = 0$ corresponds to
$r=0$.

Eqs.(\ref{uvfi0}) generalize as follows for the dominant order
\begin{eqnarray}\label{uvtefi}
u(\s,\tau)= e^{a(\s)} \left[ 1\right. &-&\left. {1\o 4} \, \gamma(\s)^4 \;
\tau^{ 4/5} +\ldots \right] , \cr \cr
v(\s,\tau)= e^{-a(\s)} \left[ 1\right. &-&\left. {1\o 4} \,
\gamma(\s)^4 \; \tau^{ 4/5}  +\ldots \right] , \cr \cr
\theta(\s,\tau)=  g(\s) &+&   \tau^{1/5}\; \mu(\s) +\ldots , \cr \cr
\phi(\s,\tau)=   f(\s) &+&  \tau^{1/5}\; \nu(\s) +\ldots .
\end{eqnarray}
Inserting eqs.(\ref{uvtefi}) in  eqs.(\ref{eqang},\ref{eqks}) and constraints
(\ref{vinks}) yields
\begin{equation}\label{munug}
  [2 \,  \gamma(\s)]^2 =\mu(\s)^2 +  \nu(\s)^2 \; \sin^2 g(\s) \; .
\end{equation}
For $g(\s) \equiv \pi/2,  \mu(\s)^2 \equiv 0$ we get back the previous
equatorial solution.

We can also find the ring solution of ref.\cite{hjile} setting $ f(\s)
 \equiv n \s, a(\s) \equiv 0, \mu(\s)= $ cte.  $g(\s)= $ cte.   and $
\nu(\s) \equiv 0$.

The coordinate $r$  vanishes here as for the equatorial solution
\begin{equation}\label{erre}
r(\s,\tau) =\gamma(\s)^2 \;  \tau^{2/5} + \ldots \; .
\end{equation}

The string solution is completely fixed once
the functions  $ f(\s), g(\s), a(\s), \mu(\s)$ and $ \nu(\s)$ are
given. These five functions are arbitrary and can be expressed in
terms of the  initial data.

Notice that $\phi$ and $\theta$ approach their limiting values with
the same exponent $1/5$ in $\tau$.

As in the equatorial case, the corrections to  the leading behaviour
appear as positive integer powers of $\tau^{2/5}$. The subdominant
leading power in $u(\s,\tau)$ and $v(\s,\tau)$ is again
$\tau^{7/5}$. We find with the help of Mathematica,
\begin{eqnarray}\label{uvmas}
u(\s,\tau)&=& e^{a(\s)} \left\{ 1\right. -   \, \gamma(\s)^4 \;
\tau^{ 4/5}  \left[ 1 + O(\tau^{2/5}) \right] \cr \cr
&-& \left.  \,\gamma(\s)^6  \;
{{ f'(\sigma)\nu(\sigma)\sin^2g(\sigma)+\mu(\sigma)}\over
{28\;  a'(\sigma)}}\; \tau^{7/5} \left[ 1 + O(\tau^{2/5}) \right] \right\}
\; , \cr \cr
v(\s,\tau)&=& e^{-a(\s)} \left\{ 1\right. -  \, \gamma(\s)^4 \;
\tau^{ 4/5}  \left[ 1 + O(\tau^{2/5}) \right]  \cr \cr
&+& \left.   \,\gamma(\s)^6  \;
{{ f'(\sigma)\nu(\sigma)\sin^2g(\sigma)+\mu(\sigma)}\over
{28 \;  a'(\sigma)}}\; \tau^{7/5} \left[ 1 + O(\tau^{2/5}) \right] \right\}
\; .
\end{eqnarray}
Using the diagonal conformal transformation (\ref{diag}), we can fix
one of the arbitrary functions among  $ f(\s), g(\s), a(\s), \mu(\s)$
and $ \nu(\s)$ keeping in mind the periodic boundary conditions:
\begin{eqnarray}
 a(\s+ 2\pi ) =  a(\s)\; , \; \nu(\s + 2\pi )= \nu(\s)\; , \; \mu(\s +
2\pi )= \mu(\s)\; , \cr \cr
\; f(\s+ 2\pi)=f(\s) + 2 n \pi \; , \;  g(\s+2\pi)= g(\s) \; \mbox{mod}\;  2
\pi \; .
\end{eqnarray}

We are left with {\bf four} arbitrary functions of $\s$. This is
precisely the number of transverse string degrees of freedom.

\section{String energy-momentum  and  invariant size near the
singularity and the near the horizon}

 The invariant string size $ds^2$ is defined using the  metric induced  on
the string world-sheet \cite{erice}:
\begin{equation}\label{intinv}
ds^2 = G_{AB}(X) \, {\dot X}^A \, {\dot X}^B \left(d\tau^2 - d\s^2\right).
\end{equation}
We give the  name {\it string size} to
\begin{equation}\label{tama}
S = -  G_{AB}(X) \, {\dot X}^A \, {\dot X}^B = {4 \o r}\, e^{-r}
{\dot u}{\dot v} -   r^2
\,  {\dot\theta}^2 - r^2 \, {\dot\phi}^2 \, \sin^2\theta .
\end{equation}
We find near the singularity at $r=0$ using eqs.(\ref{uvfi0}-\ref{r0})
\begin{eqnarray}\label{tamaO}
S &=& {{ 4\,  {a'(\s)}^2}\o { \gamma(\s)^2}} \; \tau^{ -2/5} -
{4 \o 7}\; a'(\s)^2 \; \left( 6 + 25 \; {{ a'(\s)^2}\o
{\gamma(\s)^7}} \right) + O(\tau^{2/5}) \cr \cr
&=& {{ 4\,  {a'(\s)}^2}\o r}  -
{4 \o 7}\; a'(\s)^2 \; \left( 6 + 25 \; {{ a'(\s)^2}\o
{\gamma(\s)^7}} \right) + O(r) \; .
\end{eqnarray}
The invariant string size tends then to infinite when the string falls
into the   $r=0$ singularity. This is due to the infinitely growing
gravitational forces that act there on the string.

The string stretching near  $r=0$ was observed in ref.\cite{cl} using
perturbative methods.

\bigskip

We find near the horizon using eqs.(\ref{domhor})-(\ref{rhor}) and
(\ref{tama}),
\begin{equation}\label{tamaH}
S = [p_0'(\s)]^2 + O(\tau) =  [p_0'(\s)]^2 + O(r-1)\; .
\end{equation}
 As expected, we find a finite result  since $r=1$ is a regular point of
the geometry.

\bigskip

The spacetime string energy-momentum tensor in four spacetime
dimensions is given by
\begin{eqnarray}
\sqrt{-G}~ T^{AB}(X) = \frac{1}{2\pi \alpha'} \int d\sigma d\tau
\left( {\dot X}^A {\dot X}^B -X'^A X'^B \right)
\delta^{(4)}(X - X(\sigma, \tau) ) \; .
\label{tens}
\end{eqnarray}
Notice that $X$ in eq.(\ref{tens}) is just a spacetime point whereas $
X(\sigma, \tau)$ stands for the string dynamical variables. One sees
from  the Dirac delta in eq.(\ref{tens}) that $ T^{AB}(X) $ vanishes
unless $X$ is exactly on the string world-sheet.
We shall not be interested in the detailed structure of the classical strings.
It is  more useful to integrate
the energy-momentum tensor (\ref{tens}) on a volume
that completely encloses the string as in
refs.\cite{ijm},\cite{erice}. We obtain in this way a  density
$\Theta^{AB}$.

Inside the horizon we can use  $t, \theta, \phi$
as spatial coordinates and $r$ as a coordinate time. We find,
\begin{equation}\label{Teta}
\sqrt{g_{rr}}\,\Theta^{AB}(r)  =  \frac{1}{2\pi \alpha'} \int d\s d\tau
\left( {\dot X}^A {\dot X}^B -X'^A X'^B \right)
\delta(r - r(\tau,\s ) ).
\end{equation}
where $g_{rr} = r/(1-r) > 0$.

Let us first compute the trace of the energy-momentum tensor
(\ref{tens}). We find

\begin{equation}
\sqrt{-G}~ T^A_A(X) = \frac{1}{\pi \alpha'} \int d\sigma d\tau\;
G_{AB}(X) \;  {\dot X}^A {\dot X}^B \;
\delta^{(4)}(X - X(\sigma, \tau) ) \; .
\label{traza}
\end{equation}
where we used the string constraints. Notice that the integrand is
just the (minus) string size [eq.(\ref{tama})].

We have for the black hole case:
\begin{equation}
G_{AB}(X) \;  {\dot X}^A {\dot X}^B = -{{r {\dot r}^2}\o{1-r}} + r^2
\,  {\dot\theta}^2 + r^2 \, {\dot\phi}^2 \, \sin^2\theta + {{1-r}\o
r}\, {\dot t}^2 \; .
\end{equation}
Using eqs.(\ref{uvtefi}) and (\ref{erre}) for $\tau \to 0$, we find
that each of the first three terms grows as $\tau^{-4/5}$
whereas the last term vanishes as  $\tau^{2/5}$. Moreover,  the sum
of the three terms $O(\tau^{-4/5})$ identically vanishes thanks to
eq.(\ref{munug}). This cancellation in the trace tells us that near $r=0$, the
dominant (and divergent) components $T_r^r, T_{\phi}^{\phi}$ and
$T_{\theta}^{\theta}$ yield a zero trace. This means that the string
behaves to leading order as  {\bf two}-dimensional massless
particles. This is the so-called dual to unstable behaviour \cite{erice}
(here for two spatial dimension).

For $\tau \to 0, \; r \to 0$ we can use in eq.(\ref{Teta}) the dominant
behaviours:
\begin{eqnarray}
r(\s,\tau) &=&\gamma(\s)^2 \;  \tau^{2/5} + O( \tau^{4/5}) , \cr \cr
\theta(\s,\tau)&=&  g(\s) +  \mu(\s) \, \tau^{1/5} + O(
\tau^{3/5})  , \cr \cr
\phi(\s,\tau)&=&   f(\s) +  \, \nu(\s)  \, \tau^{1/5} + O(
\tau^{3/5})  , \cr \cr
t(\s,\tau)& =& - 2 a(\s)  +
 \gamma(\s)^6 \;
{{f'(\sigma)\nu(\sigma)\sin^2g(\sigma)+\mu(\sigma)}\over
{14 \;  a'(\sigma)}}\; \tau^{7/5}
+ O\left(\tau^{9/5}\right)\; .
\end{eqnarray}
We thus find for $ r \to 0 $,
\begin{eqnarray}
2\pi\alpha'\,\Theta^{rr}(r) & = & {2\o {5\;r^2}}
\int_0^{2\pi}{\rm d}\s\, |\gamma(\s)|^5 + O({1 \o r})
\to +\infty \,,\cr\cr
2\pi\alpha'\,\Theta^{\phi\phi}(r) & = &  {1\o {10\;r^3}}
\int_0^{2\pi}{\rm d}\s\, \nu(\s)^2 \; |\gamma(\s)|^3
+ O({1 \o {r^2}}) \to +\infty\,,\cr\cr
2\pi\alpha'\,\Theta^{\theta\theta}(r) & = &  {1\o {10\;r^3}}
\int_0^{2\pi}{\rm d}\s\, \mu(\s)^2 \; |\gamma(\s)|^3 + O({1 \o
{r^2}}) \to +\infty \,,\cr\cr
2\pi\alpha'\,\Theta^{tt}(r) & = & -10\;r
\int_0^{2\pi}{\rm d}\s\,{{\left[a'(\s)\right]^2}\o{ |\gamma(\s)|^5
}}+ O(r^2)\to 0^-
\, \, .
\end{eqnarray}

We can identify the string energy with the mixed component
$-\Theta_r^r$. We define the mixed components $\Theta_A^B(r)$
by integrating $T_A^B(X)$ over the spatial volume.

This yields for $r\to 0$,
\begin{equation}\label{ener}
E \equiv -\Theta_r^r = { 1 \o {2\pi\alpha'}}\;
 {2\o {5\;r}}
\int_0^{2\pi}{\rm d}\s\, |\gamma(\s)|^5 + O(1) \to +\infty \; .
\end{equation}

The transverse pressures are defined as the mixed components
$\Theta_{\phi}^{\phi}$ and $\Theta_{\theta}^{\theta}$. They diverge
for $r\to 0 $ :
\begin{eqnarray}\label{pres}
P_{\phi} \equiv \Theta_{\phi}^{\phi} &=&  { 1 \o {2\pi\alpha'}}\;  {2\o {5\;r}}
\int_0^{2\pi}{\rm d}\s\, \nu(\s)^2 \;  \sin^2 g(\s) \; |\gamma(\s)|^5
\,\to +\infty\,,\cr\cr
P_{\theta} \equiv \Theta_{\theta}^{\theta} &=&  { 1 \o {2\pi\alpha'}}\;
{2\o {5\;r}}
\int_0^{2\pi}{\rm d}\s\, \mu(\s)^2 \; |\gamma(\s)|^5 \,\to +\infty .
\end{eqnarray}
Thus, to leading order,
$$
E = P_{\theta}+ P_{\phi} \quad \mbox{for}\quad r \to 0 \; .
$$
exhibiting a two-dimensional ultrarelativistic gas behaviour.
The tidal forces  infinitely stretch the
string near $r=0$ in effectively only two directions: $\phi$ and $\theta$.

We find for the off-diagonal components,
\begin{eqnarray}
2\pi\alpha'\,\Theta^{t r}(r) &=& {{r^{1/2}}\over 10}\; \int_0^{2\pi}{{{\rm
d}\s\,}\o{a'(\s)}}\, \gamma(\s)^4\;
\left[f'(\sigma)\nu(\sigma)\sin^2g(\sigma)+\mu(\sigma)\right]\to 0^+
\; , \cr\cr
2\pi\alpha'\,\Theta^{t\theta}(r) &=& \, {2\over 5}\; \int_0^{2\pi}{\rm
d}\s\,{{\mu(\s)}\o{a'(\s)}}\, |\gamma(\s)|^3 \;
\left[f'(\sigma)\nu(\sigma)\sin^2g(\sigma)+\mu(\sigma)\right] = O(1)\; , \cr\cr
2\pi\alpha'\,\Theta^{t\phi}(r)  &=& \, {2\over 5}\; \int_0^{2\pi}{\rm
d}\s\,{{\nu(\s)}\o{a'(\s)}}\, |\gamma(\s)|^3 \;
\left[f'(\sigma)\nu(\sigma)\sin^2g(\sigma)+\mu(\sigma)\right] = O(1)\; , \cr\cr
2\pi\alpha'\,\Theta^{r \phi }(r)  &=&\, {1\o {5\;r^{5/2}}}
\int_0^{2\pi}{\rm d}\s\, \nu(\s)\,\gamma(\s)^4 \,\to \infty , \cr\cr
2\pi\alpha'\,\Theta^{r \theta}(r)  &=&\, {1\o {5\;r^{5/2}}}
\int_0^{2\pi}{\rm d}\s\, \mu(\s)\, \gamma(\s)^4\;
\,\to \infty  , \cr\cr
2\pi\alpha'\,\Theta^{\theta\phi}(r)  &=&\, {1\o {10 \;r^3}}
\int_0^{2\pi}{\rm d}\s\,  \mu(\s)\, \nu(\s)\, |\gamma(\s)|^3\; \to \infty \; .
\end{eqnarray}

Notice that the invariant string size tends to infinity [see
eq.(\ref{tamaO})] with $4\, a'(\s)^2$ as proportionality factor. Since
$-2 a(\s)$ is the leading behaviour of $t(\s,\tau)$, this suggests us that
the string stretches  infinitely in the (spatial) $t$ direction when $r\to 0$.

As a matter of fact, infinitely growing string sizes are not observed in
cosmological spacetimes  \cite{erice,backr} for strings exhibiting radiation
 (dual to unstable) behaviour.

For particular string solutions the energy-momentum tensor and the
string size can be less singular than in the generic case.
For ring solutions \cite{hjile},  $\mu(\s) = \mu =$ constant, $g(\s) =
g =$  constant, $a(\s) = \nu(\s) = 0$,
there is no stretching and
$$
S = r \sin{g} \to 0
$$
$$
E =  P_{\theta} =  { 1 \o {\alpha'}}\;{{\mu^5}\o {80\;r}} \,\to
+\infty
\quad , \quad P_{\phi} = 0 \; .
$$
There is no string stretching but the string keeps exhibiting dual to
unstable behaviour. This is due to the balance of the tidal forces
thanks to the special symmetry  of the solution.  It behaves in this
special case as {\bf one}-dimensional massless particles for $r \to 0$.

As is easy to see, setting $\mu(\s) = 0, g(\s) = \pi/2$
all equatorial string solutions   behave as {\bf
one}-dimensional massless particles for $r \to 0$.

It must be noticed that the resolution method used here for strings in
black hole spacetimes is analogous  the expansions for $\tau\to 0$
used in ref.\cite{sv,gsv} for  strings in cosmological spacetimes.

\bigskip

The string coordinates are regular near the horizon [see eqs.
(\ref{domhor})]. Let us see that the energy-momentum tensor is also
regular  near the horizon. We use now Kruskal-Szekeres coordinates and
define $\Theta^{AB}(v)$ by integration over $u, \theta$ and $\phi$.
[For simplicity, we  consider the equatorial solution].
 We find for the energy  $\Theta^{t_K t_K}(v)$,
\begin{equation}
2\pi\alpha'\,\Theta^{t_K t_K}(v=0)= {{ e^{-1/2}}\o  4} \;
\int_0^{2\pi}{\rm d}\s\,  {{ [b_1(\s) + e \, c_1(\s)]^2 - b_0'(\s)^2}\o
{|c_1(\s)|}} = O(1)
\end{equation}
All other components of  $\Theta^{AB}(t_K)$ can be analogously computed
and  take finite values. We find for  $\Theta^{r_K r_K}(v)$,
\begin{equation}
2\pi\alpha'\,\Theta^{r_K r_K}(v=0)= {{ e^{-1/2}}\o  4} \;
\int_0^{2\pi}{\rm d}\s\,  {{ [b_1(\s) - e \, c_1(\s)]^2 - b_0'(\s)^2}\o
{|c_1(\s)|}} = O(1)
\end{equation}

For the energy-momentum trace we find
\begin{equation}
\pi\alpha'\,\Theta^A_A(t_K) = e^{1/2} \; \int_0^{2\pi}{\rm d}\s\,
[p_0(\s)']^2 \,  = O(1)
\end{equation}
We see that $\Theta^A_A(t_K)>0$ near the horizon.

\section{Strings far away from the black hole}

At large distances from the black hole ($r>>1$), the string equations
of motion and constraints become those of Minkowski spacetime,

\begin{eqnarray}\label{dalam}
 \partial_{-+}X^{A}(\s,\tau)=0~, \quad 0 \le A \le 3 , \cr
 \left[ \partial_{\pm}X^{0}(\s,\tau) \right]^2 - \sum_{j=1}^{3}
\left[ \partial_{\pm}X^{j}(\s,\tau) \right]^2 = 0 \; .
\end{eqnarray}
The solution of eqs.(\ref{dalam})   takes the customary form
\begin{equation}  \label{solM}
         X^{A}(\s,\tau)  =  q^A + 2 p^A \a' \tau + i \sqrt{\a'} \sum_{n \neq 0}
   \frac{1}{n}\{ \a^{A}_{n} \exp[in(\s - \tau)]
  + \tilde  \a^{A}_{n} \exp[-in(\s + \tau )] \}  \; ,
 \label{solm}
\end{equation}
 where  $q^A$  and  $p^A$  stand for the string center of mass position
  and momentum and  $ \a^{A}_{n}$  and $\tilde \a^{A}_{n}$  describe the right
 and left oscillator modes of the string, respectively. In order to
have $r>>1$, we must assume $\tau >> 1$ so $ |q^A + 2 p^A \a' \tau| >>
1$.
Therefore, we can get the large distance approximation  here as a
systematic expansion in inverse powers of $\tau$. As zeroth order
solution we can take:
$$
q^A + 2 p^A \a' \tau \; .
$$
This is precisely the expansion constructed in ref.\cite{negro}. We
refer to this article for strings propagating far away from the
black hole. This regime is a weak field approximation. It can be
considered as a newtonian approximation plus a post-newtonian one plus
higher orders.

\section{An Exact String Solution}

An exact solution of the string equations of motion and constraints
[eqs. (\ref{eqang},\ref{eqks}-\ref{vinks})] is given by
\begin{eqnarray}\label{trivi}
u &=& f(x_-) \quad ,  \quad v = g(x_+) \; , \cr
\phi &=& 0  \quad ,  \quad \theta = \pi/2 \; .
\end{eqnarray}
where $f(x)$ and $g(x)$ are arbitrary periodic functions of $x$ with
period $2 \pi$. Even if the world-sheet reaches the  $r=0$
singularity, it is not necessarily true that we can perform a
conformal transformation on this solution such that $\tau=0$ corresponds
to   $r=0$.

Let us see that this exact radial string solution has in fact zero
energy-momentum. Inserting eqs.(\ref{trivi}) in  eq.(\ref{tens}), we
see  immediately that
all components of $ T^{AB}(X) $ vanish except possibly for  $ T^{UV}(X) $.
For this one,
\begin{eqnarray}\label{tuv}
2\pi \alpha' \sqrt{-G}~ T^{UV}(X) &=&  \int dx_+ dx_-
\left(\partial_+ U \partial_- V + \partial_- U \partial_+ V \right)
\delta^{(4)}(X - X(\sigma, \tau) ) \;  \cr
 &=&\int dx_+ dx_- f'(x_-) g'(x_+)\delta^{(4)}(X - X(\sigma, \tau) )  \cr
&=& \delta(\phi)\delta(\theta-\pi/2)\sum_i sign[ f'(x_-^i(u)) ]\sum_j
 sign[ g'(x_+^j(v)) ]
\end{eqnarray}
where $x_-^i(u)$ and $x_+^j(v)$ are the real roots of
\begin{equation}\label{inver}
u =  f(x_-) \quad ,  \quad v = g(x_+) \; ,
\end{equation}
for given $u$ and $v$. Now, since $f(x_-)$ and $g(x_+)$ are periodic
functions, eqs.(\ref{inver}) must have an even number of solutions. In
addition, $ f'(x_-^i(u))$ and $ g'(x_+^j(v)) $ will have positive and negative
signs pairwise. Therefore, the sum over $i,j$ will vanish in eq.(\ref{tuv}):
$$
 T^{UV}(X) = 0 \; .
$$
In summary, this exact radial solution does not carry
energy-momentum. It is then a pure gauge object.

Solutions of this type have been previously discussed
\cite{njrh,barssch,bars} and found to be either pure gauge or
unphysical as they stand.
In particular, the unphysical character of these solutions is
suggested in ref.\cite{barssch}, insofar as they require (on top of
the periodicity in
$\sigma$) that the physical time coordinate be an increasing function of
$\tau$. This leads to folded string solutions, which are exact solutions
parts of whose worldsheets are
described by expressions (\ref{trivi}), but with discontinuities in the first
derivatives of the worldsheet to spacetime maps. In such a case, the
stress-energy tensor is concentrated on the folds, as follows immediately
from
our computation above when applied to the specific situation at hand, thus
leading to a system with closer resemblance to a grid than to a string
worldsheet. That is to say, the lattice introduced for ``mathematical
convenience" only \cite{barssch} has been mapped into a ``grid" with
physical
significance in spacetime itself.

\appendix
\section{}
In this appendix we give the explicit expansion of the string
coordinates $u(\s,\tau), v(\s,\tau), r(\s,\tau)$ and $\phi(\s,\tau)$
around the $r=0$ singularity for the equatorial case. We choose a
gauge on the world-sheet such that  $r=0$ corresponds to $\tau = 0$.

\begin{eqnarray} \label{uapendice}
e^{a}u & = &1
-{{{{\gamma}^4}}\over4}\tau^{4/5}-\nonumber\\ &  &
-\Bigg({{2\,{{\gamma}^6}}\over {21}}+
  {{25\,{{a'}^2}}\over {14\,{{\gamma
}^4}}}\Bigg)\tau^{6/5}
-{{ {{\gamma }^7}\,f'}\over {14\,a'}}\tau^{7/5}+\nonumber\\ &  &
+\Bigg(-{{65\,{{\gamma }^8}}\over {4704}} +
  {{250\,{{a'}^2}}\over {441\,{{\gamma }^2}}} +
  {{3125\,{{a'}^4}}\over {588\,{{\gamma }^{12}}}}\Bigg)\tau^{8/5}-\nonumber\\ &
 & -\Bigg({{ {{\gamma }^9}\,f'  }\over
    {21\,a'}} + {{25\,a'\,f'}\over
    {126\,\gamma }}\Bigg)\tau^{9/5}+\nonumber\\ & & +\Bigg({{71\,{{\gamma
}^{10}}}\over {113190}} +
  {{112975\,{{a'}^2}}\over {271656}} -
  {{18750\,{{a'}^4}}\over {3773\,{{\gamma }^{10}}}} -
  {{234375\,{{a'}^6}}\over {7546\,{{\gamma }^{20}}}} -
  {{{{\gamma }^{10}}\,{{f'}^2}}\over
    {88\,{{a'}^2}}} - {{5\,a'\,\gamma '}\over
    {3\,\gamma }} + {{5\,a''}\over 6}\Bigg)\tau^2+\nonumber\\ & &
+\Bigg(-{{173\,{{\gamma }^{11}}\,f'}\over {12936\,a'}} -
  {{650\,\gamma \,a'\,f'}\over {4851}} +
  {{1250\,{{a'}^3}\,f'}\over {1617\,{{\gamma
}^9}}}\Bigg)\tau^{11/5}+\nonumber\\
& & +\Bigg({{278263\,{{\gamma }^{12}}}\over
{395531136}} -
  {{117475\,{{\gamma }^2}\,{{a'}^2}}\over {18540522}} +
  {{85896875\,{{a'}^4}}\over {40452048\,{{\gamma }^8}}} +
  {{448750000\,{{a'}^6}}\over
    {9270261\,{{\gamma }^{18}}}} +
  {{701171875\,{{a'}^8}}\over
    {3090087\,{{\gamma }^{28}}}} -\nonumber\\
& &\qquad-
  {{225\,{{\gamma }^2}\,{{f'}^2}}\over {1001}} -
  {{12\,{{\gamma }^{12}}\,{{f'}^2}}\over
    {1001\,{{a'}^2}}} -
  {{25\,\gamma \,a'\,\gamma '}\over {21}} +
  {{625\,{{a'}^3}\,\gamma '}\over
    {63\,{{\gamma }^9}}} +
  {{5\,{{\gamma }^2}\,a''}\over {21}} -
  {{125\,{{a'}^2}\,a''}\over {63\,{{\gamma }^8}}}\Bigg)\tau^{12/5}+\nonumber\\
& & +\Bigg(-{{977\,{{\gamma }^{13}}\,f'}\over {588588\,a'}} +
  {{285125\,{{\gamma }^3}\,a'\,f'}\over
    {3531528}} - {{93125\,{{a'}^3}\,f'}\over
    {441441\,{{\gamma }^7}}} -
  {{1390625\,{{a'}^5}\,f'}\over
    {294294\,{{\gamma }^{17}}}} -
  {{{{\gamma }^{13}}\,{{f'}^3}}\over
    {1144\,{{a'}^3}}} - \nonumber\\
& & \qquad -
  {{125\,{{\gamma }^2}\,f'\,\gamma '}\over
    {126}} + {{25\,{{\gamma }^3}\,f'\,a''}\over
    {147\,a'}} + {{25\,{{\gamma }^3}\,f''}\over
    {882}}\Bigg)\tau^{13/5}+\nonumber\\
& &+\Bigg({{97897\,{{\gamma }^{14}}}\over
{576816240}} -
  {{191605625\,{{\gamma }^4}\,{{a'}^2}}\over
    {4153076928}} - {{78452875\,{{a'}^4}}\over
    {194675481\,{{\gamma }^6}}} -
  {{9069078125\,{{a'}^6}}\over
    {259567308\,{{\gamma }^{16}}}} -
  {{10747656250\,{{a'}^8}}\over
    {21630609\,{{\gamma }^{26}}}} - \nonumber\\
& & \qquad -
  {{26869140625\,{{a'}^{10}}}\over
    {14420406\,{{\gamma }^{36}}}} -
  {{3460\,{{\gamma }^4}\,{{f'}^2}}\over {63063}} -
  {{3529\,{{\gamma }^{14}}\,{{f'}^2}}\over
    {672672\,{{a'}^2}}} -
  {{21625\,{{a'}^2}\,{{f'}^2}}\over
    {42042\,{{\gamma }^6}}} -
  {{95\,{{\gamma }^3}\,a'\,\gamma '}\over {252}} + \nonumber\\
& & \qquad +
  {{1250\,{{a'}^3}\,\gamma '}\over
    {1323\,{{\gamma }^7}}} -
  {{31250\,{{a'}^5}\,\gamma '}\over
    {441\,{{\gamma }^{17}}}} -
  {{2\,{{\gamma }^2}\,{{\gamma '}^2}}\over 3} -
  {{25\,{{\gamma }^4}\,a''}\over {504}} -
  {{250\,{{a'}^2}\,a''}\over
    {1323\,{{\gamma }^6}}} + \nonumber\\
& & \qquad +
  {{6250\,{{a'}^4}\,a''}\over
    {441\,{{\gamma }^{16}}}} -
  {{{{\gamma }^3}\,\gamma ''}\over 6}\Bigg)\tau^{14/5}+\nonumber\\
& &+\Bigg({{467\,{{\gamma }^{15}}\,f'}\over
{5992896\,a'}} +
  {{256225\,{{\gamma }^5}\,a'\,f'}\over
    {9270261}} - {{50344625\,{{a'}^3}\,f'}\over
    {222486264\,{{\gamma }^5}}} +
  {{42781250\,{{a'}^5}\,f'}\over
    {9270261\,{{\gamma }^{15}}}} +
  {{213671875\,{{a'}^7}\,f'}\over
    {6180174\,{{\gamma }^{25}}}} - \nonumber\\
& & \qquad -
  {{5\,{{\gamma }^{15}}\,{{f'}^3}}\over
    {3003\,{{a'}^3}}} -
  {{295\,{{\gamma }^5}\,{{f'}^3}}\over
    {24024\,a'}} - {{28825\,{{\gamma }^4}\,f'\,
      \gamma '}\over {38808}} -
  {{55625\,{{a'}^2}\,f'\,\gamma '}\over
    {6468\,{{\gamma }^6}}} +
  {{170\,{{\gamma }^5}\,f'\,a''}\over
    {1029\,a'}} +\nonumber\\
& & \qquad + {{124625\,a'\,f'\,
      a''}\over {67914\,{{\gamma }^5}}} -
  {{4525\,{{\gamma }^5}\,f''}\over {271656}} -
  {{15625\,{{a'}^2}\,f''}\over
    {135828\,{{\gamma }^5}}}\Bigg)\tau^3 +O(\tau^{16/5})\; .\nonumber\\
\end{eqnarray}

 The expression for the coordinate  $v(\s,\tau)$ follows by changing
$a \to -a, u \to v$ in both sides of eq.(\ref{uapendice}).

We find for $r(\s,\tau)$ near $r=0$:
\begin{eqnarray} \label{rapendice}
r & = & {{\gamma }^2}\tau^{2/5}+\nonumber\\
 &  &+\Bigg(-{{{{\gamma }^4}}\over 7} +
  {{25\,{{a'}^2}}\over {7\,{{\gamma }^6}}}\Bigg)\tau^{4/5}+\nonumber\\
 &  &-\Bigg({{4\,{{\gamma }^6}}\over {147}} +
  {{1850\,{{a'}^2}}\over {441\,{{\gamma }^4}}} +
  {{2500\,{{a'}^4}}\over {147\,{{\gamma }^{14}}}}\Bigg)\tau^{6/5}+\nonumber\\
 &  &+\Bigg(
-{{89\,{{\gamma }^8}}\over {11319}} +
  {{40850\,{{a'}^2}}\over {33957\,{{\gamma }^2}}} +
  {{908125\,{{a'}^4}}\over {33957\,{{\gamma }^{12}}}} +
  {{1390625\,{{a'}^6}}\over {11319\,{{\gamma }^{22}}}} +
  {{{{\gamma }^8}\,{{f'}^2}}\over {44\,{{a'}^2}}}\Bigg)\tau^{8/5}+\nonumber\\
 &  &+\Bigg(
-{{2735\,{{\gamma }^{10}}}\over {1030029}} +
  {{90400\,{{a'}^2}}\over {3090087}} -
  {{142715000\,{{a'}^4}}\over
    {9270261\,{{\gamma }^{10}}}} -
  {{811718750\,{{a'}^6}}\over
    {3090087\,{{\gamma }^{20}}}} -
  {{1068359375\,{{a'}^8}}\over
    {1030029\,{{\gamma }^{30}}}} + \nonumber\\
 &  &\qquad+
  {{1475\,{{f'}^2}}\over {4004}} +
  {{9\,{{\gamma }^{10}}\,{{f'}^2}}\over
    {2002\,{{a'}^2}}}\Bigg)\tau^{2}+\nonumber\\
 &  &+\Bigg(
{{125\,f'\,\gamma '}\over {63}} -
  {{50\,\gamma \,f'\,a''}\over
    {147\,a'}} - {{25\,\gamma \,f''}\over {441}}\Bigg)\tau^{11/5}+\nonumber\\
 &  &+\Bigg(
-{{105436\,{{\gamma }^{12}}}\over {108153045}} -
  {{69500\,{{\gamma }^2}\,{{a'}^2}}\over {64891827}} +
  {{739998250\,{{a'}^4}}\over {194675481\,{{\gamma }^8}}} +
  {{45536000000\,{{a'}^6}}\over
    {194675481\,{{\gamma }^{18}}}} +
  {{14182812500\,{{a'}^8}}\over
    {4991679\,{{\gamma }^{28}}}} + \nonumber\\
 &  &\qquad+
  {{205929687500\,{{a'}^{10}}}\over
    {21630609\,{{\gamma }^{38}}}} -
  {{18230\,{{\gamma }^2}\,{{f'}^2}}\over {63063}} +
  {{38\,{{\gamma }^{12}}\,{{f'}^2}}\over
    {21021\,{{a'}^2}}} +
  {{8375\,{{a'}^2}\,{{f'}^2}}\over
    {84084\,{{\gamma }^8}}} +
  {{4\,{{\gamma '}^2}}\over 3} +
  {{\gamma \,\gamma ''}\over 3}\Bigg)\tau^{12/5}+\nonumber\\
 &  &+\Bigg(
-{{4175\,{{\gamma }^2}\,f'\,\gamma '}\over
    {19404}} + {{98125\,{{a'}^2}\,f'\,
      \gamma '}\over {9702\,{{\gamma }^8}}} -
  {{40\,{{\gamma }^3}\,f'\,a''}\over
    {1029\,a'}} - {{83375\,a'\,f'\,
      a''}\over {33957\,{{\gamma }^7}}} +
  {{11125\,{{\gamma }^3}\,f''}\over {135828}} + \nonumber\\
 &  &\qquad+
  {{29375\,{{a'}^2}\,f''}\over
    {67914\,{{\gamma }^7}}}\Bigg)\tau^{13/5}+O(\tau^{14/5})\; .\nonumber\\
\end{eqnarray}

We get for  the angular coordinate $\phi(\s,\tau)$,
\begin{eqnarray} \label{phiapendice}
\phi&=& f + 2\,\gamma \tau^{1/5} +\nonumber\\
 &  &+\Bigg(
{{4\,{{\gamma }^3}}\over {21}} -
  {{100\,{{a'}^2}}\over {21\,{{\gamma }^7}}}\Bigg)\tau^{3/5}+\nonumber\\
 &  &+\Bigg(
{{34\,{{\gamma }^5}}\over {735}} +
  {{940\,{{a'}^2}}\over {441\,{{\gamma }^5}}} +
  {{4250\,{{a'}^4}}\over {147\,{{\gamma }^{15}}}}\Bigg)\tau+\nonumber\\
 &  &+\Bigg(
{{164\,{{\gamma }^7}}\over {11319}} -
  {{2600\,{{a'}^2}}\over {33957\,{{\gamma }^3}}} -
  {{1037500\,{{a'}^4}}\over {33957\,{{\gamma }^{13}}}} -
  {{2562500\,{{a'}^6}}\over {11319\,{{\gamma }^{23}}}} -
  {{{{\gamma }^7}\,{{f'}^2}}\over {77\,{{a'}^2}}}\Bigg)\tau^{7/5}+\nonumber\\
 &  &+\Bigg(
{{2258\,{{\gamma }^9}}\over {441441}} -
  {{25700\,{{a'}^2}}\over {1324323\,\gamma }} +
  {{47665000\,{{a'}^4}}\over {3972969\,{{\gamma }^{11}}}} +
  {{507625000\,{{a'}^6}}\over
    {1324323\,{{\gamma }^{21}}}} +
  {{882031250\,{{a'}^8}}\over {441441\,{{\gamma }^{31}}}} -
  {{500\,{{f'}^2}}\over {9009\,\gamma }} - \nonumber\\
 &  &\qquad-
  {{19\,{{\gamma }^9}\,{{f'}^2}}\over
    {3003\,{{a'}^2}}}\Bigg)\tau^{9/5}+\nonumber\\
 &  &+\Bigg(
{{20\,f'\,\gamma '}\over {63\,\gamma }} +
  {{20\,f'\,a''}\over {147\,a'}} +
  {{265\,f''}\over {882}}\Bigg)\tau^{2}+\nonumber\\
 &  &+\Bigg(
{{29912\,{{\gamma }^{11}}}\over {15450435}} -
  {{48200\,\gamma \,{{a'}^2}}\over {9270261}} -
  {{42554000\,{{a'}^4}}\over {27810783\,{{\gamma }^9}}} -
  {{7690000000\,{{a'}^6}}\over
    {27810783\,{{\gamma }^{19}}}} -
  {{303125000\,{{a'}^8}}\over {64827\,{{\gamma }^{29}}}} - \nonumber\\
 &  &\qquad-
  {{58421875000\,{{a'}^{10}}}\over
    {3090087\,{{\gamma }^{39}}}} +
  {{745\,\gamma \,{{f'}^2}}\over {63063}} -
  {{64\,{{\gamma }^{11}}\,{{f'}^2}}\over
    {21021\,{{a'}^2}}} +
  {{7250\,{{a'}^2}\,{{f'}^2}}\over
    {21021\,{{\gamma }^9}}} +
  {{4\,{{\gamma '}^2}}\over {3\,\gamma }} +
  {{\gamma ''}\over 3}\Bigg)\tau^{11/5}+\nonumber\\
 &  &+\Bigg(
-{{5275\,\gamma \,f'\,\gamma '}\over {19404}} -
  {{26875\,{{a'}^2}\,f'\,\gamma '}\over
    {3234\,{{\gamma }^9}}} +
  {{190\,{{\gamma }^2}\,f'\,a''}\over
    {3087\,a'}} + {{235250\,a'\,f'\,
      a''}\over {101871\,{{\gamma }^8}}} -
  {{325\,{{\gamma }^2}\,f''}\over {45276}} - \nonumber\\
 &  &\qquad-
  {{131875\,{{a'}^2}\,f''}\over
    {203742\,{{\gamma }^8}}}\Bigg)\tau^{12/5}+O(\tau^{13/5})\; .\nonumber\\
\end{eqnarray}

We found analogous formulas for the general  (non-equatorial) case with
the help of Mathematica.

\end{document}